\documentclass[10pt,aps,prl,reprint,superscriptaddress,
amsfonts,amssymb,amsmath,floatfix]{revtex4-1}
\usepackage{microtype}
\usepackage{lmodern}
\usepackage[utf8]{inputenc}
\usepackage[T1]{fontenc}
\usepackage{graphicx}
\usepackage{epstopdf}%
\usepackage{xcolor}

\begin{document}

\title{Overdoping graphene beyond the van Hove singularity}

\author{Philipp Rosenzweig}
\email{p.rosenzweig@fkf.mpg.de}
\author{Hrag Karakachian}
\affiliation{Max-Planck-Institut für Festkörperforschung, Heisenbergstraße 1, 70569 Stuttgart, Germany}
\author{Dmitry Marchenko}
\affiliation{Helmholtz-Zentrum Berlin für Materialien und Energie, Elektronenspeicherring BESSY II, Albert-Einstein-Straße 15, 12489 Berlin, Germany}
\author{Kathrin Küster}
\author{Ulrich Starke}
\affiliation{Max-Planck-Institut für Festkörperforschung, Heisenbergstraße 1, 70569 Stuttgart, Germany}

\date{\today}

\begin{abstract}
At very high doping levels the van Hove singularity in the $\pi^*$ band of graphene becomes occupied and exotic ground states possibly emerge, driven by many-body interactions. Employing a combination of ytterbium intercalation and potassium adsorption, we $n$ dope epitaxial graphene on silicon carbide past the $\pi^*$ van Hove singularity, up to a charge carrier density of $5.5\times10^{14}$ cm$^{-2}$. This regime marks the unambiguous completion of a Lifshitz transition in which the Fermi surface topology has evolved from two electron pockets into a giant hole pocket. Angle-resolved photoelectron spectroscopy confirms these changes to be driven by electronic structure renormalizations rather than a rigid band shift. Our results open up the previously unreachable beyond-van-Hove regime in the phase diagram of epitaxial graphene, thereby accessing an unexplored landscape of potential exotic phases in this prototype two-dimensional material.
\end{abstract}

\maketitle

\begin{figure*}[t]
	\includegraphics[width=1.0\textwidth]{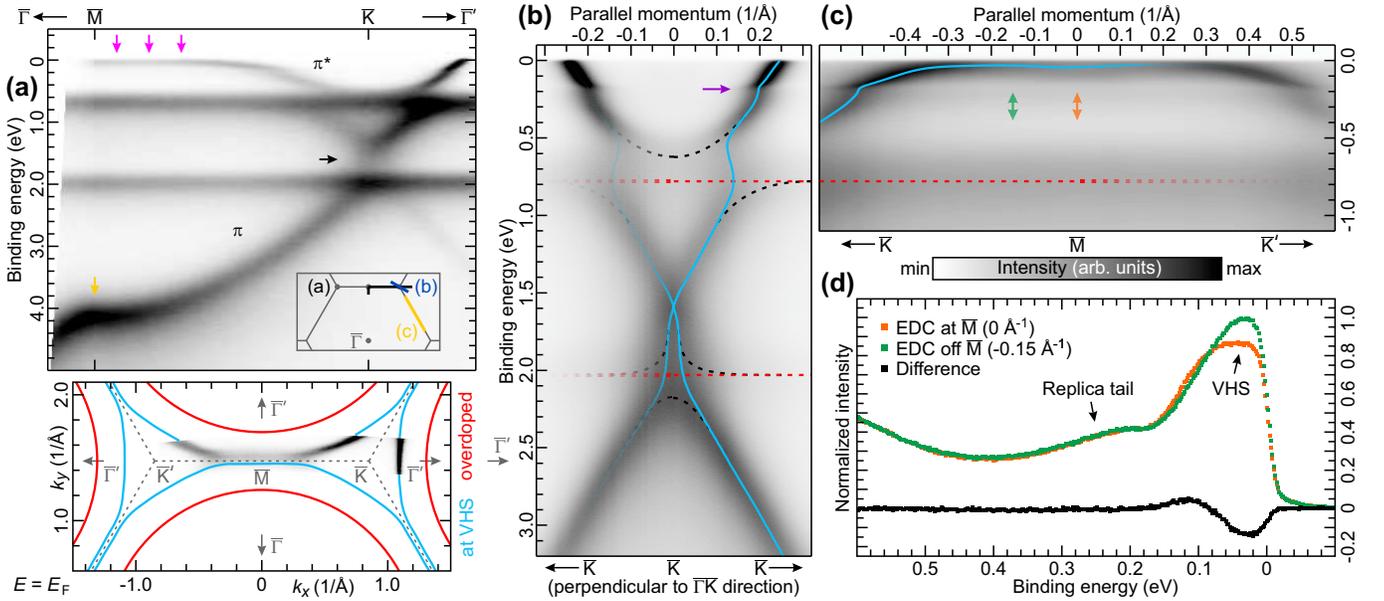}
	\caption{(a) Dispersion along $\overline{\Gamma\mathrm{MK}}$ (top) and FS (bottom) of Yb-intercalated graphene ($h\nu=80$ eV). The inset shows the orientations of the datasets (a)--(c). Solid curves in the bottom panel indicate the FS hole pocket at van Hove filling (blue) and beyond (red, exagerated for clarity). (b), (c) High-resolution energy-momentum cuts ($h\nu=40$ eV) recorded (b) at $\overline{\mathrm{K}}$ perpendicular to $\overline{\Gamma\mathrm{K}}$ and (c) along $\overline{\mathrm{KMK}}'$. Solid blue curves track the spectral maxima of the graphene $\pi$ bands which hybridize (dashed black guides to the eye) with the Yb $4f$ states (dashed red lines) near $\overline{\mathrm{K}}$. (d) EDCs extracted from (c) at $\overline{\mathrm{M}}$ ($0$ {\AA}$^{-1}$, orange) and $0.15$ {\AA}$^{-1}$ away (green) together with their difference curve (black).}
	\label{fig1}
\end{figure*}

Graphene, a honeycomb monolayer of carbon atoms, has been extensively studied in view of its two-dimensional (2D) massless Dirac fermions \cite{berger2004,novoselov2005,geim2007,katsnelson2007,neto2009}. Recently, the focus has shifted towards inducing correlated phases in this prototype 2D material \cite{wehling2011,kotov2012}. The strategy is to increase the density of states at the Fermi level $E_\mathrm{F}$, thus boosting many-body interactions and instabilities towards the pursued ground states. Possible routes are band structure engineering in twisted graphene bilayers \cite{cao2018a,cao2018b,kerelsky2019,lu2019} or flat band formation in epitaxial bilayer graphene \cite{marchenko2018}. Alternatively, excessive $n$ doping of quasi-freestanding monolayer graphene (QFMLG) on SiC pushes the $\pi^*$ van Hove singularity (VHS) into the vicinity of $E_\mathrm{F}$ \cite{mcchesney2010,rosenzweig2019,link2019}. There, many-body interactions warp the pointlike VHS into a flat band pinned to $E_\mathrm{F}$ along the $\overline{\mathrm{KMK}}'$ Brillouin zone (BZ) border: an extended van Hove scenario, reminiscent of high-$T_c$ superconductors \cite{khodel1990,gofron1994,lu1996,irkhin2002,storey2007,yudin2014}. Upon reaching the VHS, the Fermi surface (FS) undergoes a Lifshitz transition \cite{lifshitz1960,volovik2017} whereby its topology changes from two electron pockets into a single hole pocket \cite{mcchesney2010,rosenzweig2019,link2019}. Concomitantly, theory predicts various ordered ground states such as chiral superconductivity or charge density wave \cite{herbut2006,honerkamp2008,nandkishore2012,kiesel2012,blackschaffer2014,makogon2011,raghu2008}. The exact filling factors of the $\pi^*$ band at which the individual phases may stabilize are however hard to pinpoint in calculations, further complicated by the severe band structure renormalizations which have not been taken into account until recently \cite{link2019}. Experimentally, a tunable $n$-type doping near van Hove filling is therefore desirable. While the carrier density can be reduced away from the VHS \cite{rosenzweig2019}, overdoping has not yet been unambiguously confirmed and remains an important task at the basis of exploring the phase diagram of highly-doped graphene. As more and more electrons are transferred onto QFMLG, the $\pi^*$ hole pocket in the FS will naturally shrink [Fig.~\ref{fig1}(a) bottom], providing direct evidence for overdoping past the Lifshitz transition.

Here, we $n$ dope epitaxial graphene on SiC to beyond the $\pi^*$ VHS. Carrier densities of up to $5.5\times10^{14}$ cm$^{-2}$ are achieved via ytterbium intercalation of the $(6\sqrt{3}\times6\sqrt{3})\mathrm{R}30^\circ$ carbon buffer layer and subsequent potassium deposition. Angle-resolved photoelectron spectroscopy (ARPES) reveals a shrinking $\pi^*$ hole pocket, unambiguously confirming the completion of the Lifshitz transition. The latter is driven by renormalizing $\pi$ bands rather than a rigid band shift, demonstrating the extensive influence of many-body interactions at extreme doping levels. Upon overdoping a previously inaccessible regime is reached in the phase diagram of graphene, where exotic ordered ground states might emerge \cite{herbut2006,honerkamp2008,nandkishore2012,kiesel2012,blackschaffer2014,makogon2011,raghu2008}.

The $(6\sqrt{3}\times6\sqrt{3})\mathrm{R}30^\circ$ carbon buffer layer was grown on 6H-SiC(0001) following \textcite{emtsev2009}. Its transformation into QFMLG via intercalation of ytterbium at a sample temperature of $250^\circ$C is detailed elsewhere \cite{rosenzweig2019}. The QFMLG samples were transferred to the synchrotron facility in an ultrahigh vacuum suitcase (Ferrovac GmbH) at a pressure $<1\times10^{-9}$ mbar, ensuring sample stability such that no additional annealing was necessary. ARPES was performed at the $1^2$ endstation of the UE112 beamline at BESSY II, Helmholtz-Zentrum Berlin~\cite{helmholtz2018} using linearly polarized synchrotron radiation and a hemispherical analyzer (Scienta R8000) with a 2D electron detector. The angular resolution was $0.1^\circ$ while the maximally achieved energy resolution of the entire setup (beamline and analyzer) was $20$ meV. During all experiments the sample was kept at $\approx 20$ K and the base pressure was $<5\times10^{-10}$ mbar. Potassium was deposited \emph{in situ} on the cold sample from a commercial alkali metal dispenser (SAES Getters; operated at $5.6$ A, $1.3$ V).

We first set a basis for our overdoping study by presenting high-resolution, low-temperature ARPES of the extended van Hove scenario in pristine Yb-intercalated graphene. The top panel of Fig.~\ref{fig1}(a) gives an overview of the dispersion along $\overline{\Gamma\mathrm{MK}}$, probed at a photon energy of $h\nu=80$ eV. Nondispersive Yb $4f$ states at $\approx 0.8$ and $\approx 2.0$ eV below $E_\mathrm{F}$ are superimposed on the graphene $\pi$ bands. Their conical dispersions near $\overline{\mathrm{K}}$ and $\overline{\mathrm{K}}'$ (black arrow) are connected by the lower $\pi$ band through a saddle point $\approx 4.1$ eV below $E_\mathrm{F}$ at $\overline{\mathrm{M}}$ (yellow arrow). Due to strong $n$ doping the $\pi^*$ band becomes occupied to a large extent and its $\overline{\mathrm{M}}$-point VHS reaches $E_\mathrm{F}$. Contrary to the lower saddle point, the $\pi^*$ VHS is warped into a flat band pinned to $E_\mathrm{F}$ over a wide ($\approx 0.6$ {\AA}$^{-1}$) momentum range along $\overline{\mathrm{KMK}}'$ (magenta arrows), generating a FS hole pocket around $\overline{\Gamma}$ [Fig.~\ref{fig1}(a) bottom]---the extended van Hove scenario \cite{mcchesney2010,rosenzweig2019,link2019,khodel1990,gofron1994,lu1996,irkhin2002,storey2007,yudin2014}. In a single-particle picture \cite{neto2009} the Dirac point $(E_\mathrm{D})$ should lie halfway between the lower and upper saddle points, i.e., about $2.0$ eV below $E_\mathrm{F}$. However, $\vert E_\mathrm{F}-E_\mathrm{D}\vert\approx 1.6$ eV in our data. Instead of a purely rigid shift, the $\pi^*$ band bends down and reduces its bandwidth at $\overline{\mathrm{M}}$ by more than $1$ eV (cf.\ Refs.\ \cite{neto2009,bostwick2007}), therefore reaching the VHS at an electron density on the order of $3\times10^{14}$ cm$^{-2}$ \cite{rosenzweig2019} (cf.\ below).

Figure \ref{fig1}(b) shows a cut through the Dirac cone at $\overline{\mathrm{K}}$ perpendicular to $\overline{\Gamma\mathrm{K}}$ ($h\nu=40$ eV). The solid blue curves track the spectral maxima of the momentum distribution curves (MDCs). We identify the Dirac point (the branch crossing) at a binding energy of $E_\mathrm{D}=1.58\pm0.02$ eV, confirming with higher accuracy its previous determination \cite{rosenzweig2019}. The conical dispersion of graphene is distorted around the Yb $4f$ states (dashed red lines). This becomes most obvious near the $4f_{7/2}$ level at a binding energy of $0.78 \pm 0.01$ eV where the group velocity of the $\pi^*$ band seems to change sign. This unphysical MDC-derived dispersion strongly suggests an anti-crossing-type $\pi$--$4f$ hybridization (dashed black guides to the eye)---a scenario supported by band structure calculations for Yb sandwiched in between a freestanding graphene bilayer \cite{hwang2014}. The prominent kink $0.19\pm0.01$ eV below $E_\mathrm{F}$ (purple arrow) can be ascribed to electron-phonon interaction \cite{bostwick2007,tse2007,forti2011}. From the slope of the real part of the spectral function near $E_\mathrm{F}$ \cite{*[{Using a parabolic bare band and the algorithm of: }][{.}] pletikosic2012} we estimate a total coupling strength $\lambda \approx 0.4$, compatible with previous studies of highly $n$-doped graphene on SiC \cite{hwang2014,ludbrook2015}.

Figure \ref{fig1}(c) tracks the $\pi^*$ dispersion along $\overline{\mathrm{KMK}}'$. The solid blue curve is fitted to the MDC (for binding energies $>0.1$ eV) and energy distribution curve (EDC, for parallel momenta $\vert k_\parallel\vert<0.45$ {\AA}$^{-1}$) maxima. Interestingly, the flat band seems to disperse very slightly ($<20$ meV) through a local minimum at $\overline{\mathrm{M}}$. This tiny dip might be a consequence of the changing EDC lineshape from the edge towards the center of the flat band [Fig.\ \ref{fig1}(d)]. The VHS signature in the EDC at $\overline{\mathrm{M}}$ ($0$ {\AA}$^{-1}$, orange) is slightly broader, less intense and peaks into a $\approx 40$ meV wide plateau as compared to the sharper EDC at $-0.15$ {\AA}$^{-1}$ (green) (cf.\ black difference curve). Linewidth broadening of the flat band along $\overline{\mathrm{KMK}}'$, suggesting a varying quasiparticle scattering rate, has not yet been reported in similar systems \cite{mcchesney2010,link2019}. Naively, one would expect the scattering probability to peak right at $\overline{\mathrm{M}}$ if the $\pi^*$ VHS were still pointlike. In turn, our data suggest that the $\overline{\mathrm{M}}$ point's singular character is partially retained despite the extension of the VHS in $k$ space. The decreasing intensity of the flat band towards $\overline{\mathrm{M}}$ could be explained by spectral weight redistribution upon VHS extension and linewidth broadening. While being also reminiscent of pseudogap formation as encountered in other correlated materials \cite{valla2006,vishik2010} only temperature-dependent studies at ultimate resolution might shed light on this issue. The flat band at $E_\mathrm{F}$ is accompanied by a replica $\approx 0.2$ eV below, which seems to emerge out of the kinks at $\pm 0.5$ {\AA}$^{-1}$ in Fig.\ \ref{fig1}(c) and whose width and intensity do not vary significantly along $\overline{\mathrm{KMK}}'$ [Fig.\ \ref{fig1}(d)]. The same tail exists in the VHS regime of alkali-metal-doped \cite{mcchesney2010} as well as Gd-intercalated graphene \cite{link2019} and was ascribed to polaron formation resulting from the coupling to optical phonons of graphene at $\overline{\Gamma}$. A coupling constant $\lambda_{\overline{\Gamma}}\approx 1$ to this specific mode was estimated in Ref.\ \cite{link2019}, which remains a reasonable value in the present case considering the similar intensity ratio ($\approx 2.2$) between the VHS peak and the plateau $0.2$ eV below. These resemblances suggest that the feature is intrinsic to the extended van Hove scenario of graphene and, apart from the necessary charge transfer, independent of the respective dopants. Apparently, $\lambda_{\overline{\Gamma}}$ is strongly enhanced compared to graphene outside the VHS regime \cite{fedorov2014}. This is compatible with the hole pockets of adjacent BZs running parallel and very close to each other over a wide range along $\overline{\mathrm{KMK}}'$ [Fig.~\ref{fig1}(a) bottom] such that FS nesting will promote resonant coupling to the aforementioned zone-center phonons \cite{link2019}. While the flat band formation (not its replica) was linked to spin fluctuations within graphene \cite{link2019}, charged impurity scattering---promoted via a disordered intercalant \cite{rosenzweig2019}---could also account for $\pi^*$ band flattening, pronounced kinks along $\overline{\mathrm{KMK}}'$ and a flat band replica near the kink energy \cite{kaasbjerg2019, *kaasbjerg2019b}. The observed features might therefore emerge from a combination of these different mechanisms.

\begin{figure*}[t]
	\includegraphics[width=1.0\textwidth]{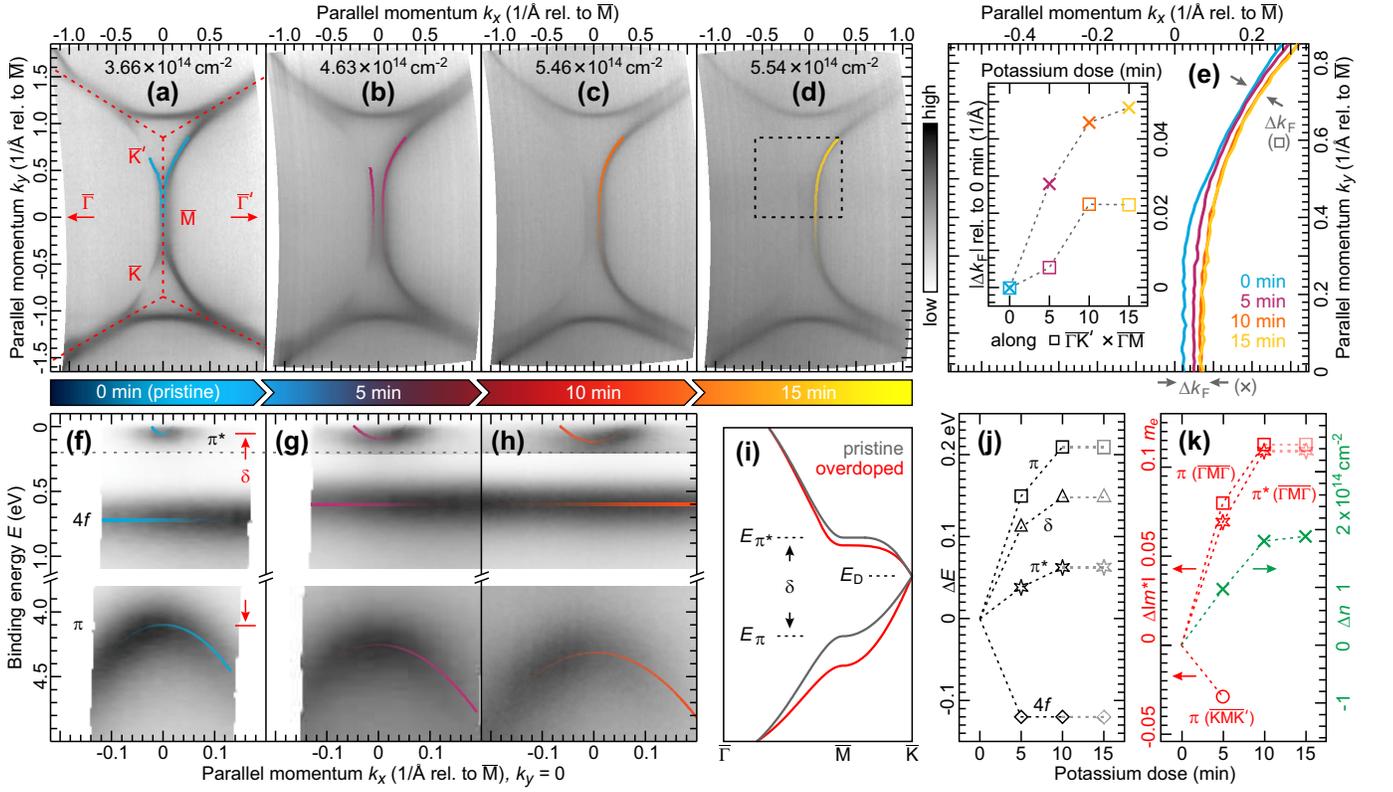}
	\caption{(a)--(d) Fermi surfaces (logarithmic grayscale) and (f)--(h) energy-momentum cuts along $\overline{\Gamma\mathrm{M}\Gamma}'$ ($h\nu=80$ eV) for increasing potassium dose (color-encoded). Solid curves are fits to the data and grayscales differ between and within the individual panels to enhance the individual spectral features. (e) FS contours extracted from (a)--(d) within the dashed rectangle overlayed in (d). The inset shows the evolution $\vert \Delta k_\mathrm{F}\vert$ of the Fermi wavevector along $\overline{\Gamma\mathrm{M}}$ and $\overline{\Gamma\mathrm{K}}'$  relative to the pristine sample as a function of potassium dose. (i) Schematized $\pi$-band renormalizations upon overdoping (exagerated for clarity). (j) Relative binding energy changes $\Delta E$ of $\pi$, $\pi^*$, Yb $4f_{7/2}$ and the $\pi$--$\pi^*$ gap $\delta$ at $\overline{\mathrm{M}}$ with potassium dose. (k) Relative changes $\Delta$ in the charge carrier density $n$ (green) and the absolute effective masses $\vert m^*\vert$ of $\pi$, $\pi^*$ at $\overline{\mathrm{M}}$ (red). Transparent data points for 15 min exposure are extrapolated based on the saturation of $n$.}
	\label{fig2}
\end{figure*}

Figure \ref{fig2}(a) displays the FS of Yb-intercalated graphene, covering large portions of the first and adjacent BZs ($h\nu=80$ eV). Entailed by the extended van Hove scenario, the two electron pockets centered at $\overline{\mathrm{K}}$ and $\overline{\mathrm{K}}'$ demonstrate strong triangular warping and merge into a giant hole pocket around $\overline{\Gamma}$ via a Lifshitz transition \cite{lifshitz1960,volovik2017}. The FS is quantified by fitting the MDCs in $k_x$ direction for $\vert k_y\vert<0.85$ {\AA}$^{-1}$. From the contour around $\overline{\Gamma}'$ in the repeated BZ [blue curves in Figs.\ \ref{fig2}(a), (e)] and following Luttinger's theorem \cite{luttinger1960,luttinger1960a} we infer an electron density of $n=(3.66\pm0.40)\times10^{14}$ cm$^{-2}$ \footnote{The vertical $\overline{\mathrm{KMK}}'$ BZ border does not perfectly coincide with $k_x=0$ in experiment. The fitted contours [Figs.\ \ref{fig2}(a)--\ref{fig2}(e)] are corrected for these deviations considering the FS symmetry relative to $\overline{\mathrm{KMK}}'$}. Being separated by only $\approx 0.02$ {\AA}$^{-1}$ from the $\overline{\mathrm{KMK}}'$ border, the individual hole-like contours of adjacent BZs cannot be unambiguously discerned, i.e., the VHS remains pinned to $E_\mathrm{F}$ and the system is held right at the Lifshitz transition. Upon additional charge transfer onto QFMLG (i.e., overdoping), the completion of the Lifshitz transition could be directly visualized by a shrinking hole pocket [Fig.~\ref{fig1}(a) bottom]. We thus performed sequential, 5-min-long deposition cycles of potassium, a well-known $n$-type dopant for epitaxial graphene \cite{ohta2006}. After the first dosage the distance between the hole contour and the $\overline{\mathrm{KMK}}'$ line has increased by a factor of $2.5$ to $\approx 0.05$ {\AA}$^{-1}$ [purple curves in Figs.\ \ref{fig2}(b), (e)] and the FS fit confirms an elevated carrier density of $n=(4.63\pm0.20)\times10^{14}$ cm$^{-2}$. Note also the suppression of the $\pi^*$ band inside the first BZ ($k_x<0$) due to matrix-element effects (dark corridor) \cite{gierz2011} and the deteriorating signal-to-noise ratio following potassium adsorption. After 10 min of exposure the hole pockets have shrunk to such extent that their individual contours are clearly distinguishable across the $\overline{\mathrm{KMK}}'$ borders and the Lifshitz transition has unambiguously concluded at a charge carrier density of $n=(5.46\pm0.20)\times10^{14}$ cm$^{-2}$ [Fig.\ \ref{fig2}(c)]. Finally, during a third deposition cycle, the FS does no longer change significantly, indicating a saturation of doping within the error range [Fig.\ \ref{fig2}(d), $n=(5.54\pm0.20)\times10^{14}$ cm$^{-2}$]. Figure \ref{fig2}(e) displays a zoomed-in view of the fitted FS contours of Figs.\ \ref{fig2}(a)--\ref{fig2}(d) used to derive the carrier densities. For pristine Yb-intercalated graphene (blue) the giant hole pocket assumes the shape of a flattened circle as the VHS extends along $\overline{\mathrm{KMK}}'$. With increasing potassium dose its contour becomes more and more circular (purple, orange) until saturation sets in (yellow). The inset tracks the absolute changes $\vert \Delta k_\mathrm{F}\vert$ of the Fermi wavevector along $\overline{\Gamma\mathrm{M}}$ and $\overline{\Gamma\mathrm{K}}'$ relative to the pristine sample. While the initial evolution is restricted to the extended-VHS sector and the $\overline{\Gamma\mathrm{K}}'$ direction remains basically unaffected, the hole pocket shrinks rather isotropically during the second deposition cycle. This suggests that surpassing the Fermi level pinning of the VHS is indeed the critical step in achieving the desired overdoping.

The energy-momentum cuts of Figs.\ \ref{fig2}(f)--\ref{fig2}(h) reveal the evolution of the electronic structure in the vicinity of $\overline{\mathrm{M}}$ towards $\overline{\Gamma}$ and $\overline{\Gamma}'$ with increasing potassium dose \footnote{Corresponding data for 15 min exposure not available}. The solid curves are parabolic fits to the spectral maxima and highlight the upward (downward) dispersion of the $\pi^*$ ($\pi$) band perpendicular to $\overline{\mathrm{KMK}}'$ \footnote{In Figs.\ \ref{fig2}(f)--\ref{fig2}(h), the fits to the $\pi^*$ band adopt fixed $E_\mathrm{F}$ crossings according to the FS fits of Figs.\ \ref{fig2}(a)--\ref{fig2}(c).}. The changes in binding energy $E$ and absolute effective mass $\vert m^*\vert$ at $\overline{\mathrm{M}}$ of the individual bands are plotted relative to the pristine sample in Figs.\ \ref{fig2}(j) and \ref{fig2}(k), respectively. With increasing potassium exposure the data quality degrades and the spectral weight of the $\pi$ bands is progressively quenched. In combination with the decreased energy resolution of $\approx 0.1$ eV, errors of up to $50$ \% (in particular for $m^*$) can result and we do not aim at an overly quantitative analysis. Yet our data support a consistent picture of $\pi$-band renormalizations in overdoped graphene [Fig.\ \ref{fig2}(i)], which will hopefully stimulate further theoretical modelling. First we note that the binding energy of the Yb $4f_{7/2}$ state decreases upon potassium exposure, qualitatively matching the previously reported increase when reducing the $n$ doping away from the VHS \cite{rosenzweig2019}. In order to reach the extended van Hove scenario of Figs.\ \ref{fig2}(a) and \ref{fig2}(f), the $\pi^*$ band had to bend down considerably whereas the $\pi$ bandwidth ($E_\pi-E_\mathrm{D}\approx 2.5$ eV) is only slightly lower than for intrinsic monolayer graphene on SiC ($\approx 2.8$ eV) \cite{bostwick2007}. As such the $\pi$--$\pi^*$ bandgap at $\overline{\mathrm{M}}$, $\delta\approx 4.1$ eV, has decreased compared to the tight-binding description relevant for lower doping levels. While both $E_\pi$ and $E_{\pi^*}$ increase with potassium doping [Fig.\ \ref{fig2}(j)], the $\pi$ band is much more affected, contrary to when QFMLG is initially doped up to the VHS. In consequence of this nonrigid band warping beyond a simple shift of $E_\mathrm{F}$, $\delta$ has already expanded by more than $0.1$ eV following the first deposition cycle. That is, the trend of a reduced $\pi$--$\pi^*$ gap with increasing $n$ doping \cite{mcchesney2010,rosenzweig2019,link2019} reverses once QFMLG is overdoped past the VHS. The system thus counteracts the imposed overdoping and tries to keep the $\pi^*$ VHS close to $E_\mathrm{F}$ by pushing the $\pi$ bands away from each other at $\overline{\mathrm{M}}$ [Fig.\ \ref{fig2}(i)]. This scenario is substantiated by the apparent increase in $\vert m^*\vert$ (i.e., decrease in curvature) along $\overline{\Gamma\mathrm{M}\Gamma}'$ for both bands [Fig.\ \ref{fig2}(k)]. In the orthogonal $\overline{\mathrm{KMK}}'$ direction, the absolute effective mass $\vert m^*\vert$ of the $\pi$ band at $\overline{\mathrm{M}}$ decreases, suggesting that the Dirac point energy $E_\mathrm{D}$ does not move rigidly with $E_\pi$ [Fig.\ \ref{fig2}(i)]. At this point, a substantial downshift of $E_\mathrm{D}$ together with the entire $\pi$-band system can also be excluded due to the basically constant size of the FS hole pocket in $\overline{\Gamma\mathrm{K}}'$ direction [Fig.\ \ref{fig2}(e)]. $E_\mathrm{D}$ therefore appears to act as a relatively inert reference level of the renormalizing $\pi$-band structure near van Hove filling, consistent with QFMLG underdoped to below the VHS \cite{rosenzweig2019}. Although $n$ increases linearly over the first two deposition cycles [Fig.\ \ref{fig2}(k)], the $\pi$ bands and their effective masses along $\overline{\Gamma\mathrm{M}\Gamma}'$ seem to renormalize considerably less during the second exposure \footnote{Data along $\overline{\mathrm{KMK}}'$ not available}. This could be explained by the fading influence of the VHS Fermi-level pinning, now that it has clearly been surpassed. In parallel, the FS evolves more isotropically [Fig.\ \ref{fig2}(e)] indicating a transition to a doping-induced shift of the entire $\pi$-band system instead of band warping which dominates across the Lifshitz transition. As the observed band structure renormalizations cannot be described in a single-particle picture \cite{link2019,ulstrup2016} our experiments highlight the influence of nonlocal many-body effects in graphene under extreme $n$-type (over)doping.

In summary, epitaxial graphene on SiC is $n$ doped past the extended van Hove singularity, up to a carrier density of $5.5\times10^{14}$ cm$^{-2}$, through a combination of ytterbium intercalation and potassium adsorption. Angle-resolved photoelectron spectroscopy confirms the overdoping by probing a shrinking hole pocket in the Fermi surface. The completion of the associated Lifshitz transition is driven by electronic structure renormalizations rather than a rigid band shift, exemplified by an increasing energy gap between the upper and lower $\pi$ bands at $\overline{\mathrm{M}}$. Our results open up the previously inaccessible beyond-van-Hove regime in the phase diagram of graphene where a variety of exotic ground states may stabilize.
\begin{acknowledgments}
We would like to thank Helmholtz-Zentrum Berlin for the allocation of synchrotron radiation beamtime under proposals 182-07119-ST and 191-08299-ST/R. We are grateful to Kristen Kaasbjerg for providing valuable information on charged impurity scattering in doped graphene. This work was supported by the Deutsche Forschungsgemeinschaft (DFG) through Sta315/9-1.
\end{acknowledgments}

\end{document}